\documentstyle[aps]{revtex}

\begin{document}
\title{Carrier relaxation due to electron-electron interaction in coupled double
quantum well structures}
\author{Marcos R.S. Tavares$^{1,2}$, G.-Q Hai$^{2}$ and S. Das Sarma$^{1}$}
\address{$^{1}$Department of Physics, University of Maryland, College Park, MD\\
20742-4111, USA}
\address{$^{2}$Instituto de F\'{i}sica de S\~{a}o Carlos, Universidade de S\~{a}o\\
Paulo, S\~{a}o Carlos, SP 13560-970, Brazil}
\maketitle

\begin{abstract}
We calculate the electron-electron interaction induced energy-dependent
inelastic carrier relaxation rate in doped semiconductor coupled double
quantum well nanostructures within the two subband approximation at zero
temperature. In particular, we calculate, using many-body theory, the
imaginary part of the full self-energy matrix by expanding in the
dynamically RPA screened Coulomb interaction, obtaining the intrasubband and
intersubband electron relaxation rates in the ground and excited subbands as
a function of electron energy. We separate out the single particle and the
collective excitation contributions, and comment on the effects of
structural asymmetry in the quantum well on the relaxation rate. Effects of
dynamical screening and Fermi statistics are automatically included in our
many body formalism rather than being incorporated in an {\it ad-hoc} manner
as one must do in the Boltzman theory.
\end{abstract}


\section{\protect\smallskip Introduction}

\smallskip Electron-electron interaction induced carrier relaxation is an
important inelastic scattering process in low-dimensional semiconductors
nanostructures. It is often (particularly in situations where LO\ phonon
emission is energetically prohibited because the excited electrons do not
have enough energy) the most dominant relaxation process in semiconductor
quantum wells and wires, and is therefore of considerable fundamental and
practical importance. Band gap engineering has led to the possibility of
fabricating tunable far infrared quantum well cascade lasers (QCL) and
efficient quantum well infrared photodetectors (QWIP) where inelastic
carrier relaxation via electron-electron interaction is a crucial (perhaps
even decisive) process in determining device operation and feasibility \cite
{1}. For QCL and QWIP operations it is the intersubband inelastic relaxation
which turns out to be the primary rate limiting scattering process. For
other proposed devices, such as the planar hot electron transistors or
related 2D high speed devices, intrasubband relaxation is the important
process. A thorough quantitative understanding of intra- and inter-subband
relaxation due to electron-electron interaction is therefore important for
the successful realization of these devices. In addition to this practical
technological motivation arising from QCL, QWIP, and other proposed
band-engineered quantum well devices, there is also obvious fundamental
reason for studying inelastic Coulomb scattering in 2D quantum well systems.
Inelastic electron-electron scattering determines the 2D quasiparticle
spectral width, as determined for example in tunneling measurements, through
the imaginary part of the electron self-energy function \cite{2}.

In this article we use a many-body approach in calculating the inelastic
relaxation rate of 2D electrons confined in GaAs-Al$_{x}$Ga$_{1-x}$ As
semiconductor quantum well structures. Our work is a multisubband
generalization of the earlier work \cite{3} by Jalabert and Das Sarma who
considered only intrasubband relaxation within a single subband model. We
consider both intra and intersubband relaxation in the two lowest subbands,
and consider both single well and coupled double well structures. An
additional important issue addressed in our work is the effect of structural
asymmetry in the quantum well on the relaxation rate. This is in fact a
potentially significant factor in the fabrication of QCL and QWIP structures
since asymmetry could lead to the opening of new electron-electron
interaction channels in the inelastic intersubband relaxation as we discuss
below in this article.

The central quantity we calculate in this work, within the leading order
dynamically screened Coulomb interaction expansion (the so-called GW
approximation in the multisubband situation), is the imaginary part of
electronic on-shell self-energy matrix, $M$, in the quantum well subband
index ($i,j,$ etc.). The subband self-energy in the multisubband situation
is, in general, off-diagonal reflecting the breaking of the translational
invariance along the growth ({\it z}) direction (we take the {\it x-y} plane
to be the 2D plane with all wave vectors in this paper being 2D wave vectors
in the {\it x-y} plane). The off-diagonal self-energy, $%
\mathop{\rm Im}%
\left( M_{ij}\right) $, incorporates in an intrinsic many-body manner the
possibility of electron-electron interaction induced intersubband scattering
(both virtual and real) of carriers. We believe that in the doped situation
of our interest, where the quantum well subbands are occupied by many
electrons, the many-body self-energy approach is the only reasonable
technique in calculating the inelastic carrier relaxation rate in contrast
to the Boltzman equation approach where the scattering rates usually
calculated using Fermi's golden rule. This is because the dynamical
screening inherent in the many electron system, which affects the calculated
inelastic scattering rates in profound and highly nontrivial way, is
automatically incorporated in our many-body GW expansion whereas inclusion
of dynamical screening in Fermi's golden rule type formula is problematic
and can only be done heuristically by replacing the bare interaction by a
screened interaction in an {\it ad-hoc} manner.

Our theory, as mentioned above, is based on the so-called GW self-energy
approximation \cite{3,4} where the electron self-energy $M$ is obtained in a
leading order expansion of the dynamically screened Coulomb interaction $%
W\equiv V^{s}$, where the superscript $s$ denotes dynamical screening of the
bare electron-electron interaction matrix $V$ in the multisubband situation.
We use the RPA to obtain the dynamical screened interaction $V^{s}$, i.e. $%
V^{s}\equiv \varepsilon ^{-1}V$, with $\varepsilon \equiv 1-V\Pi $, where $%
\Pi $ is the leading-order (i.e. noninteracting) electron polarizability
matrix. We also approximate the electron Green's function\thinspace $G$ by
the noninteracting Green's function $G^{0}$ making our formal expression for
the self-energy matrix to be 
\begin{equation}
M\sim \int G^{0}V^{s},  \label{1}
\end{equation}
where the integral involves integrating over all internal momentum and
energy variables as well as summing over all internal subband indices (and
spin). Putting the subband (matrix) indices explicitly in Eq. (\ref{1}), we
get 
\begin{equation}
\mathop{\rm Im}%
M_{ij}=%
\mathop{\rm Im}%
\sum_{lm}\int G_{lm}^{0}V_{ilmj}^{s}.  \label{2}
\end{equation}
We note, however, that $G^{0},$ being the noninteracting Green's function,
is necessarily diagonal in subband indices (i.e. an electron cannot suffer
an intersubband scattering in the absence of interaction); 
\begin{equation}
G_{lm}^{0}\sim G_{ll}^{0}\delta _{lm}.  \label{3}
\end{equation}
Then, Eq. (2) becomes 
\begin{equation}
\mathop{\rm Im}%
M_{ij}=\sum_{l}\int 
\mathop{\rm Im}%
\left[ G_{ll}^{0}V_{illj}^{s}\right] ,  \label{4}
\end{equation}
with 
\begin{equation}
V_{illj}^{s}=\left( \varepsilon ^{-1}V\right) _{illj}.  \label{5}
\end{equation}
Equations (\ref{4}) and (\ref{5}) are the central formal equations we use in
our theory to obtain the inelastic relaxation time $\tau $, remembering that
the scattering rate $\Gamma $ and the relaxation time $\tau $ are connected
by 
\begin{equation}
\tau =\frac{\hbar }{2\Gamma },  \label{6}
\end{equation}
where, 
\begin{equation}
\Gamma =\left| 
\mathop{\rm Im}%
M\right| .  \label{7}
\end{equation}

In Eqs. (\ref{6}) and (\ref{7}), $\Gamma =\left| 
\mathop{\rm Im}%
M\right| $ is calculated on-shell, i.e., the quasiparticle self-energy
defines $\Gamma $. To demonstrate how Eq. (\ref{4}) may, in principle,
differ from the simplistic Fermi's golden rule approach we consider the
specific two subband model of interest to us in this paper. Then $i,j,l=1,2$
with only subband 1, the ground subband, and the subband 2, the first
excited subband being considered in the theory assuming all other subbands
to be substantially higher in energy making negligible contributions to the
self-energies of the lowest two subbands. We also assume the square well
structure to be symmetric, so that parity is a good quantum number in the
problem which makes all ''off-diagonal'' interaction matrix elements vanish 
\cite{5} by virtue of parity conservation with the only non-zero elements of 
$V^{s}$ being $V_{1111}^{s},V_{2222}^{s}$, $V_{1212}^{s},$ and $V_{1122}^{s}$
(note that $V_{1221}^{s}=V_{2112}^{s}$ and $V_{1122}^{s}=V_{2211}^{s}\,$by
symmetry). In this situation Eq. (\ref{4}) implies that 
\begin{equation}
\mathop{\rm Im}%
M_{12}=%
\mathop{\rm Im}%
M_{21}=0,  \label{8}
\end{equation}
and, 
\begin{equation}
\mathop{\rm Im}%
M_{22}\sim \int 
\mathop{\rm Im}%
\left[ G_{11}^{0}V_{2112}^{s}+G_{22}^{0}V_{2222}^{s}\right] .  \label{9}
\end{equation}
We note that the dynamically screened interaction matrix element $%
V_{1122}^{s}$ is not explicitly present in Eq. (\ref{9}). On the other hand,
a naive Fermi's golden rule approach will explicitly include such a $%
V_{1122}^{s}$ term, as indeed has been done in the literature \cite{6},
because it seems to arise from the direct Coulomb interaction $V_{1122}$
between an electron in subband 1 and an electron in subband 2 without any
intersubband scattering. We mention, however, that dynamical screening of $%
V_{2222}$ produces an effective $V_{2211}$ term in our theory since
dynamical screening proceeds through virtual creation of electron-hole pairs.

We have calculated the energy dependent inelastic relaxation rate at $T=0$
for a two-subband (1 and 2) GaAs-Al$_{x}$Ga$_{1-x}$As quantum well system
with a total electron density $N_{e}=2\times 10^{11}$ cm$^{-2}$ for the
following five distinct situations. (i) Two coupled symmetric quantum wells
of width 150 \AA $\,$each with interwell tunneling induced by a tunnelling
barrier of height 228 $\,$meV and width 30 \AA . Here the lowest two
subbands are the so-called symmetric (bonding) and antisymmetric
(antibonding) levels with energies $E_{1}=$ 15.35 meV and $E_{2}=$ 17.03
meV, respectively. The third level $E_{3}$= 60.53 meV is sufficiently high
to be ignored ($E_{F1}=E_{F}-E_{1}=$ 4.28 meV; $E_{F2}$ $=E_{F}-E_{2}=$ 2.61
meV), with both subbands 1 and 2 occupied by carriers. These results are
presented in Section III.A below. (ii) Two coupled asymmetric quantum wells
with interwell tunneling (the same as in (i) above), with one well of width
150 \AA\ and the other of width 140 \AA\ $\,$leading to $E_{1}=$ 15.93$\,\ $%
meV and $E_{2}=$ 18.55 meV ($E_{F1}=E_{F}-E_{1}=$ 4.75 meV; $E_{F2}$ $%
=E_{F}-E_{2}=$ 2.13 meV). Again, the next excited subband $E_{3}=$ 62.86 meV
is high enough to be ignored. These results are presented in Section III.A
below. (iii) Two coupled identical symmetric quantum wells of width 150 \AA\
each with no interwell tunneling (i.e. the interwell barrier is taken to be
infinity) and with a barrier width of 30 \AA . Here, $E_{1}=E_{2}$ = 23.87
meV (this degeneracy arises because the two wells are identical and there is
no tunneling), $E_{F1}=E_{F2}$ $=E_{F}-E_{1}=E_{F}-E_{2}=$ 3.44 meV, and the
next subband $E_{3}$ $=$ 96 meV is sufficiently high in energy to be
neglected. These results are presented in Section III.B below. (iv) The same
as in the last case with no interwell tunneling but an asymmetric situation
with the two wells being different. One with a width of 150 \AA \thinspace\
and the other a width of 142.4 \AA\ so that the subband Fermi energies $%
E_{F1}=E_{F}-E_{1}=$ 4.75 meV and $E_{F2}$ $=E_{F}-E_{2}=$ 2.13 meV, which
are the same as in (ii) above. In this situation $E_{1}$ = 23.87 meV, $%
E_{2}= $ 26.49 meV (again $E_{3}$ can be neglected). Furthermore, to keep
the average distance between the two electron layers the same as in (ii), we
choose a barrier of width 28.8 \AA . These results are presented in Section
III.B below in comparison with those in (ii). (v) A single symmetric quantum
well of width 300 \AA\ and a barrier height 228$\,$\ meV, which leads to the
lowest two subbands at $E_{1}=$ 4.88 meV, $E_{2}=$ 19.51 meV, and the Fermi
energy $E_{F1}=E_{F}-E_{1}=$ 6.88 meV (with $E_{F}$ $<E_{2},$ so that the
second subband is empty). In this situation, the next excited subband, $%
E_{3}=$ 43.74 meV, is high enough in energy to be neglected. These results
are present in Section III.C below. Our reason for studying the five
different classes of systems described above is that we are interested in
understanding the effects of interwell tunneling and structural asymmetry on
the electron relaxation rate. In particular, asymmetry breaks parity
conservation making the off-diagonal matrix elements of Coulomb interaction
(e.g. $%
V_{1112},V_{1121},V_{1211},V_{2111},V_{2221},V_{2212},V_{2122},V_{1222}$
\thinspace all of which are zero in the symmetric situation) nonzero,
leading to new inelastic relaxation channels not present in symmetric
structures. For the sake of brevity we present results for a single
representative carrier density and well parameters in each of the five
cases. Our theory could be easily generalized to obtain finite temperature
relaxation rates. Note that our goal here is to provide a qualitative
understanding of how various physical parameters affect Coulomb scattering
rates in 2D quantum wells.

The plan of this article is the following. In Sec. II we present a brief
theory with working formulae; in Sec. III we provide our numerical results
and discussions; we conclude in Sec. IV with a summary.

\section{Theory}

Our basic theory is outlined in the Introduction, where the formal
expression for the self-energies to be calculated were given. Our central
GW-RPA expression \cite{7} for the self-energy can be explicitly written out
by using the noninteracting subband Green's function 
\begin{equation}
G_{ij}^{0}(\omega ,k)=\delta _{ij}\left[ \omega -E_{i}(k)+E_{F}\right] ^{-1},
\label{11}
\end{equation}
where $\omega $ is a complex frequency $(\hbar =1)$, and $%
E_{i}(k)=E_{i}+k^{2}/2m^{\ast }$ is the noninteracting subband one-electron
energy dispersion. Using Eq. (\ref{11}) in Eqs. (\ref{4}) and (\ref{5}), and
carrying out the internal frequency integration, and taking the imaginary
part after the on-shell analytic continuation, we get 
\[
\mathop{\rm Im}%
M_{ij}(k)=\frac{1}{(2\pi )^{2}}\sum_{l}\int d^{2}{\bf q}%
\mathop{\rm Im}%
\left[ V_{illj}^{s}({\bf q},\xi _{l}({\bf k}+{\bf q})-\xi _{i}({\bf k})%
\right] \times 
\]
\begin{equation}
\left\{ \theta \left( \xi _{i}({\bf k})-\xi _{l}({\bf k}+{\bf q})\right)
-\theta \left( -\xi _{l}({\bf k}+{\bf q})\right) \right\} ,  \label{12}
\end{equation}
where the on-shell subband energy $\xi _{i}({\bf k})$ is given by 
\begin{equation}
\xi _{i}({\bf k})=E_{i}({\bf k})-E_{F},  \label{13}
\end{equation}
and $\theta \left( x\right) =0$ $(1),$ for $x<(>)$ $0,$ is the usual
Heaviside theta function. The dynamically screened Coulomb interaction is
given by (see Eq. \ref{5}) 
\begin{equation}
V_{ijlm}^{s}=(\varepsilon ^{-1}V)_{ijlm},  \label{14}
\end{equation}
with the multisubband RPA approximation \cite{7} defined by the dielectric
matrix 
\begin{equation}
\varepsilon _{ijlm}=(1-V_{ijlm}\Pi _{lm}^{0}),  \label{15}
\end{equation}
where $V_{ijlm}$ is the bare Coulomb interaction matrix element in the
subband representation, and $\Pi _{ij}^{0}$, the noninteracting
polarizability, is given by 
\begin{equation}
\Pi _{ij}^{0}({\bf k},\omega )=-2\int \frac{d^{2}{\bf q}}{(2\pi )^{2}}\frac{%
f_{i}({\bf k}+{\bf q})-f_{i}({\bf k})}{\omega -E_{i}({\bf k}+{\bf q})+E_{j}(%
{\bf k})},  \label{16}
\end{equation}
where $f_{i}({\bf k})$ is the Fermi distribution function in the $i$th
subband. In this paper, we take the impurity scattering induced background
broadening $\gamma $ as being a small phenomenological damping parameter
which equivalent to be working in the clean limit. We are therefore
restricting ourselves to high mobility quantum wells with small impurity
scattering induced level broadening.

Using Eqs. (\ref{12}-\ref{16}) it is straightforward to calculate the
imaginary part of the on-shell self-energy. For the sake of completeness, we
show below the detailed expressions for $%
\mathop{\rm Im}%
M_{ij}$ in the GW approximation for the two-subband model: 
\begin{equation}
\mathop{\rm Im}%
M_{11}(k)=\sigma _{1111}(k)+\sigma _{1221}(k),  \label{17}
\end{equation}
\begin{equation}
\mathop{\rm Im}%
M_{12}(k)=\sigma _{1112}(k)+\sigma _{1222}(k),  \label{18}
\end{equation}
\begin{equation}
\mathop{\rm Im}%
M_{21}(k)=\sigma _{2111}(k)+\sigma _{2221}(k),  \label{19}
\end{equation}
and 
\begin{equation}
\mathop{\rm Im}%
M_{22}(k)=\sigma _{2112}(k)+\sigma _{2222}(k).  \label{20}
\end{equation}
Here, 
\begin{equation}
\sigma _{1111}(k)=\frac{1}{(2\pi )^{2}}\int d^{2}{\bf q}\left\{ 
\mathop{\rm Im}%
\left[ V_{1111}^{s}({\bf q},A)\right] \left[ \theta \left( -A\right) -\theta
\left( -\xi _{1}({\bf k}+{\bf q})\right) \right] \right\} ,  \label{21}
\end{equation}
\begin{equation}
\sigma _{1221}(k)=\frac{1}{(2\pi )^{2}}\int d^{2}{\bf q}\left\{ 
\mathop{\rm Im}%
\left[ V_{1221}^{s}({\bf q},A+\omega _{0})\right] \left[ \theta \left(
-A-\omega _{0}\right) -\theta \left( -\xi _{2}({\bf k}+{\bf q})\right) %
\right] \right\} ,  \label{22}
\end{equation}
\begin{equation}
\sigma _{1112}(k)=\frac{1}{(2\pi )^{2}}\int d^{2}{\bf q}\left\{ 
\mathop{\rm Im}%
\left[ V_{1112}^{s}({\bf q},A)\right] \left[ \theta \left( -A\right) -\theta
\left( -\xi _{1}({\bf k}+{\bf q})\right) \right] \right\} ,  \label{23}
\end{equation}
\begin{equation}
\sigma _{1222}(k)=\frac{1}{(2\pi )^{2}}\int d^{2}{\bf q}\left\{ 
\mathop{\rm Im}%
\left[ V_{1222}^{s}({\bf q},A+\omega _{0})\right] \left[ \theta \left(
-A-\omega _{0}\right) -\theta \left( -\xi _{2}({\bf k}+{\bf q})\right) %
\right] \right\} ,  \label{24}
\end{equation}
\begin{equation}
\sigma _{2111}(k)=\frac{1}{(2\pi )^{2}}\int d^{2}{\bf q}\left\{ 
\mathop{\rm Im}%
\left[ V_{2111}^{s}({\bf q},A-\omega _{0})\right] \left[ \theta \left(
-A+\omega _{0}\right) -\theta \left( -\xi _{1}({\bf k}+{\bf q})\right) %
\right] \right\} ,  \label{25}
\end{equation}
\begin{equation}
\sigma _{2221}(k)=\frac{1}{(2\pi )^{2}}\int d^{2}{\bf q}\left\{ 
\mathop{\rm Im}%
\left[ V_{2221}^{s}({\bf q},A)\right] \left[ \theta \left( -A\right) -\theta
\left( -\xi _{2}({\bf k}+{\bf q})\right) \right] \right\} ,  \label{26}
\end{equation}
\begin{equation}
\sigma _{2112}(k)=\frac{1}{(2\pi )^{2}}\int d^{2}{\bf q}\left\{ 
\mathop{\rm Im}%
\left[ V_{2112}^{s}({\bf q},A-\omega _{0})\right] \left[ \theta \left(
-A+\omega _{0}\right) -\theta \left( -\xi _{1}({\bf k}+{\bf q})\right) %
\right] \right\} ,  \label{27}
\end{equation}
and 
\begin{equation}
\sigma _{2222}(k)=\frac{1}{(2\pi )^{2}}\int d^{2}{\bf q}\left\{ 
\mathop{\rm Im}%
\left[ V_{2222}^{s}({\bf q},A)\right] \left[ \theta \left( -A\right) -\theta
\left( -\xi _{2}({\bf k}+{\bf q})\right) \right] \right\} ,  \label{28}
\end{equation}
where $\omega _{0}=E_{2}-E_{1}\,$ is the subband energy difference and $%
A\equiv A({\bf q},{\bf k})=(2kq\cos \eta +q^{2})/2m^{\ast }$ with $\eta $
being the angle between ${\bf k}$ and ${\bf q}$; and $m^{\ast }=0.07m_{e}$
being the GaAs conduction band electron effective mass. Now, we define the 
{\em total} inelastic Coulomb scattering rate for an electron with
wavevector $k$ ( i.e. an energy of $k^{2}/2m^{\ast }$ with respect to the
subband bottom) in the subband 1 and 2 as 
\begin{equation}
\sigma _{1}(k)=%
\mathop{\rm Im}%
M_{11}(k)+%
\mathop{\rm Im}%
M_{12}(k)  \label{29}
\end{equation}
and 
\begin{equation}
\sigma _{2}(k)=%
\mathop{\rm Im}%
M_{21}(k)+%
\mathop{\rm Im}%
M_{22}(k).  \label{30}
\end{equation}
It is important to realize that the screened potentials $V_{ijlm}^{s}$ for $%
j\neq l$ do not appear in Eqs. (\ref{21}-\ref{28}) and, consequently, do not
explicitly contribute to scattering rate. They are implicitly induced in the
theory through dynamical screening \cite{8} as discussed before.
Furthermore, all screened interactions $V_{ijlm}^{s}$ involved in the Eqs. (%
\ref{21}-\ref{28}) are obtained from the relation between the bare
electron-electron potential \cite{7} 
\[
V_{ijlm}(q)=\frac{2\pi e^{2}}{q\varepsilon _{0}}\int dz\int dz^{\prime }\phi
_{i}(z)\phi _{j}(z)e^{-q\left| z-z^{\prime }\right| }\phi _{l}(z^{\prime
})\phi _{m}(z^{\prime })
\]
and the inverse matrix of the dynamical dielectric function $\varepsilon
_{ijlm}(q,\omega )\,$ (see Eqs. (\ref{14}) and (\ref{15}), where the indices 
$i,j,l,\,$and $m=1,2$). These bare Coulomb potentials $V_{ijlm}(q)$ are
calculated here by using both the one-electron wavefunction $\phi _{i}(z)$
and the subband energy $E_{i}$ obtained through the numerical solution of
the Schr\"{o}dinger equation in the $z$ direction for the specific quantum
well confinement potential. Furthermore, the potential $V_{ijlm}(q)$ can be
separated into intra- and intersubband terms, and understood as follows: (i)
Intralayer (intrasubband) interactions $V_{1111}(q)=V_{A},$ $%
V_{2222}(q)=V_{B},$ and $V_{1122}(q)=V_{2211}(q)=V_{C}$ representing\ those
scattering events which the electrons remain in their original well
(subband); (ii) Interlayer (intersubband) interactions $%
V_{1212}(q)=V_{2121}(q)=V_{1221}(q)=V_{2112}(q)=V_{D}$ representing
scattering in which both electrons change their well (subband) indices; and
(iii) Intra-interwell (subband) interactions$\
V_{1112}(q)=V_{1121}(q)=V_{1211}(q)=V_{2111}(q)=V_{J}$ and $%
V_{2212}(q)=V_{2221}(q)=V_{1222}(q)=V_{2122}(q)=$ $V_{H}$ indicating the
scattering in which only one of the electrons suffers the interwell
(intersubband) transition.

\smallskip For each wave-vector $k,$ the two dimensional ${\bf q}$ integrals
in Eqs. (\ref{21}-\ref{28}) are performed within the planes determined
through the variables ${\bf q}$ and $A$ in the screened interactions $%
V_{ijlm}^{s}$. The integration domains of $q$ and $\eta $ (in $A$) variables
are restricted by the two theta functions appearing in the integrals in Eqs.
(\ref{21}-\ref{28}). The integrals involving $V_{1111}^{s}$ and $%
V_{1112}^{s} $ are performed within the planes formed by those regions in
the $q$ space where 
\begin{equation}
\theta \left[ -A\right] -\theta \left[ -\xi _{1}({\bf k}+{\bf q})\right]
\neq 0;  \label{31}
\end{equation}
while the integrals involving $V_{1221}^{s}$ and $V_{1222}^{s}$ are
calculated within the planes defined by 
\begin{equation}
\theta \left[ -A-\omega _{0}\right] -\theta \left[ -\xi _{2}({\bf k}+{\bf q})%
\right] \neq 0.  \label{32}
\end{equation}
In the same way, the integrals involving $V_{2222}^{s}$ and $V_{2221}^{s}$
are performed within the planes defined by 
\begin{equation}
\theta \left[ -A\right] -\theta \left[ -\xi _{2}({\bf k}+{\bf q})\right]
\neq 0;  \label{33}
\end{equation}
and, finally, for $V_{2112}^{s}$ and $V_{2111}^{s}$the integrating plane is
defined by 
\begin{equation}
\theta \left[ -A+\omega _{0}\right] -\theta \left[ -\xi _{1}({\bf k}+{\bf q})%
\right] \neq 0.  \label{34}
\end{equation}
The inelastic scattering rates in the Eqs. (\ref{21}-\ref{28}) vanish
outside each corresponding integrating plane, which means that the momentum
and energy conservation cannot be simultaneously obeyed for such values of ($%
{\bf k},{\bf k}+{\bf q}$), and therefore no Coulomb scattering is allowed
there. It is easy to see that these integrals are non-vanishing only if the
corresponding integrating plane contains either some part of the
single-particle excitation continuum or some branch representing the
collective excitations (plasmons) in the 2-D $q$ plane. This is of course
expected since a finite scattering rate must involve real excitations, which
in this case are single-particle and collective plasmon excitations.

\section{Numerical results and discussions}

\subsection{Coulomb Coupled Bilayers With Interwell Tunneling}

We consider first two coupled symmetric identical quantum wells of same
width $W_{1}=W_{2}=$150 \AA\ with an interwell tunneling induced by a
barrier of height 228$\,$meV and width 30 \AA . The total electron density $%
N_{e}=n_{1}+n_{2}=2\times 10^{11}$ cm$^{-2}$ in all structures studied in
this paper, with $n_{1}$ and $n_{2}$ being the density in the subband 1 and
2, respectively. For these sample parameters, the Fermi wavevectors in the
first and second subband are $k_{F1}^{sy}\simeq $ 0.88 $\times 10^{6}$ cm$%
^{-1}$ and $k_{F2}^{sy}$ $\simeq $ 0.69 $\times 10^{6}$ cm$^{-1}$,
respectively (the superscript $sy$ stands for symmetric). Here, both
subbands 1 and 2 (symmetric and antisymmetric, respectively) are occupied by
carriers with $n_{1}\simeq 1.23\times 10^{11}$ cm$^{-2}$ and $n_{2}\simeq
0.77\times 10^{11}$ cm$^{-2}$.

\smallskip The plasmon dispersion relation is determined by the roots of the
determinantal equation $\det \left| \varepsilon _{ijlm}(q,\omega )\right| =0$%
, which, after some algebra, can be rewritten as 
\[
\varepsilon _{intra}\varepsilon _{inter}-\left[ (1-V_{A}\Pi
_{11}^{0})V_{H}^{2}\Pi _{22}^{0}+\left( 1-V_{B}\Pi _{22}^{0}\right)
V_{J}^{2}\Pi _{11}^{0}\right. - 
\]
\begin{equation}
\left. 2V_{C}V_{J}V_{H}\Pi _{11}^{0}\Pi _{22}^{0}\left( \Pi _{12}^{0}+\Pi
_{21}^{0}\right) \right] =0,  \label{35}
\end{equation}
where 
\begin{equation}
\varepsilon _{intra}=\left( 1-V_{A}\Pi _{11}^{0}\right) \left( 1-V_{B}\Pi
_{22}^{0}\right) -V_{C}^{2}\Pi _{11}^{0}\Pi _{22}^{0}  \label{36}
\end{equation}
and 
\begin{equation}
\varepsilon _{inter}=\ 1-V_{D}\left( \Pi _{12}^{0}+\Pi _{21}^{0}\right) .
\label{37}
\end{equation}
For notational simplicity, we do not explicitly write the energy and wave
vector dependence in Eqs. (\ref{35}-\ref{37}). For the present symmetric
situation, the unscreened Coulomb potential $V_{J}=V_{H}=0$ by virtue of
parity symmetry, because the wavefunctions $\phi _{1}(z)$ and $\phi _{2}(z)$
are symmetric and antisymmetric functions of $z$, respectively. According to
the Eq. (\ref{35}), therefore, the plasmons dispersion relation in our
symmetric bilayer structure are determined by the roots of the equation $%
\varepsilon _{intra}\varepsilon _{inter}=0$, i.e. either $\varepsilon
_{intra}=0$ corresponding to the 2D intrasubband plasmons, or $\varepsilon
_{inter}=0$ corresponding to the intersubband plasmons.

There are four roots of $\varepsilon _{intra}=0$. Two of them are shown in
Fig. 1(a) by the solid lines indicating the intrasubband plasmon modes $%
(1,1) $ and $(2,2)$. Notice that, for each solid line, there is a
corresponding dashed line which is also the root of the same equation always
lying in the corresponding single-particle excitation continuum. It is well
known that the plasmon modes indicated by the dashed lines inside the
single-particle continua are strongly Landau damped by single-particle
excitations and will be ignored in the following discussion. Furthermore,
the intersubband plasmon mode $(1,2)$ comes from the roots of $\varepsilon
_{inter}=0.$ The wavefunctions $\phi _{1}(z)$ and $\phi _{2}(z)$ are
schematically shown in the inset by the solid and dot lines, respectively.
Notice that, one is always able to separate the intra- and inter-subband \
plasmon modes in structures which are invariant under space inversion. In
addition, the intrasubband plasmons are not Landau damped by intersubband
single-particle excitations and vice versa in symmetric bilayer systems. The
single-particle continua SPE$_{11}$ (intrasubband SPE) and SPE$_{12}$
(intersubband SPE) in Fig. 1(a) are those regions where $%
\mathop{\rm Im}%
\left\{ \Pi _{11}^{0}(q,\omega )\right\} \neq 0$ and $%
\mathop{\rm Im}%
\left\{ \Pi _{12}^{0}(q,\omega )\right\} \neq 0$, respectively. For the sake
of simplicity, we will not indicate the continuum SPE$_{22}$ in this paper
because it lies totally inside the continuum SPE$_{11}$. Moreover, we claim
that the plasmon mode $(2,2)$ should be strongly damped by single-particle
excitations in the SPE$_{11}$ continuum and will also be ignored in the
following qualitative scattering rate discussion. Our numerical results of
course include all contributions as obtained by evaluating the 2D integrals
in Eqs. (\ref{21}-\ref{28}).

Fig. 1(b) shows the same plasmon dispersion relation as in Fig. 1(a) but now
in a two coupled {\em asymmetric} quantum wells with an interwell tunneling.
Here, one well is of width 150 \AA\ and the other is of width 140 \AA . For
these parameters, the Fermi wave vector in the subband 1 and 2 are $%
k_{F1}^{a}\simeq $ 0.93 $\times 10^{6}$ cm$^{-1}$ and $k_{F2}^{a}\simeq $
0.62 $\times 10^{6}$ cm$^{-1}$, $\,$respectively (the superscript $a$ stands
for asymmetric). Both subbands are occupied with $n_{1}\simeq $ 1.37 $\times
10^{11}$ cm$^{-2}$ and $n_{2}\simeq $ 0.63 $\times 10^{11}$ cm$^{-2}$. In
this asymmetric situation, the plasmon modes are obtained directly from the
roots of Eq. (\ref{35}). We show in Fig. 1(b) all these roots. We mention
that it does not make sense naming the solid lines as pure intra- or
inter-subband plasmon modes because the structural asymmetry leads to a
strong coupling (or mixing) between them, and this intra-inter mode coupling
eliminates the simplicity of Fig. 1(a). The solid line that is of finite
frequency as $q\rightarrow 0$ in Fig. 1(b) is the intersubband-like plasmon
mode $(1,2)$. This mode enters the continuum SPE$_{12}$ at $q\simeq $ 0.42 $%
\times 10^{6}$ cm$^{-1}$ and should be, in principle, Landau damped. For
small values of $q$, we find the same number of roots as in the symmetric
situation. The interactions $V_{J}$ and $V_{H}$ are finite in the asymmetric
case and are responsible for the strong mixing between the intrasubband-like
plasmon mode $(1,1)\,$and the intersubband-like mode $(1,2)$ around $q\cong $
0.18$\times 10^{6}$ cm$^{-1}$. Moreover, when the asymmetry is introduced,
the depolarization shift (i.e. the shift of the intersubband plasmon from
the subband energy separation $E_{21}$) in the intersubband-like plasmon $%
(1,2)$ at $q=0$ decreases. We point out that these roots of Eq. (\ref{34})
do not provide a complete description of the plasmon modes in asymmetric
bilayer structures. A detailed theoretical calculation of the dynamical
structure factor giving the plasmon spectral weight provides a complete
picture of the collective mode spectra and can be obtained using our
multisubband theory.

Having studied the plasmon dispersion relations we now investigate in
Fig.2(a) the corresponding total inelastic Coulomb scattering rate $\sigma
_{1}(k)$ (thick-solid line) and $\sigma _{2}(k)$ (thick-dashed line) of fast
electrons in the subband 1 and 2, respectively, as a function of wavevector $%
k$ in our symmetric bilayer structure. The symbols on the thin lines
identify the contributions to $\sigma _{1}(k)$ and $\sigma _{2}(k)$ coming
from the emission of single-particle and collective excitations
individually. The dynamically screened Coulomb interaction components
entering in Eqs. (\ref{21}-\ref{28}) can be calculated from Eq. (\ref{14})
and (\ref{15}). After some algebra, we get 
\begin{equation}
V_{1111}^{s}=\frac{V_{A}\left( 1-V_{B}\Pi _{22}^{0}\right) +V_{C}^{2}\Pi
_{22}^{0}}{\varepsilon _{intra}},  \label{V1111}
\end{equation}
\begin{equation}
V_{2222}^{s}=\frac{V_{B}\left( 1-V_{A}\Pi _{11}^{0}\right) +V_{C}^{2}\Pi
_{11}^{0}}{\varepsilon _{intra}},  \label{V2222}
\end{equation}
and 
\begin{equation}
V_{1221}^{s}=V_{2112}^{s}=\frac{V_{D}}{\varepsilon _{inter}}.  \label{V1212s}
\end{equation}
For the symmetric well case the off-diagonal components of the Coulomb
potential all vanish by parity: $%
V_{1112}^{s}=V_{1121}^{s}=V_{1211}^{s}=V_{2111}^{s}=V_{2221}^{s}=V_{2212}^{s}=V_{2122}^{s}=V_{1222}^{s}=0 
$ because $V_{J}=V_{H}=0$ for symmetric systems. Therefore, according to
Eqs.(\ref{21}-\ref{30}), the total inelastic scattering rates in the subband
1 and 2 are 
\begin{equation}
\sigma _{1}(k)=\sigma _{1111}+\sigma _{1221}  \label{gam1}
\end{equation}
and 
\begin{equation}
\sigma _{2}(k)=\sigma _{2222}+\sigma _{2112},  \label{gam2}
\end{equation}
respectively. The terms $\sigma _{1111},$ $\sigma _{1221}$, $\sigma _{2222}$
and $\sigma _{2112}$ involve integrations of the interactions $V_{1111}^{s},$
$V_{1221}^{s},$ $V_{2222}^{s}$ and $V_{2112}^{s}$, respectively; in Eqs. (%
\ref{21}), (\ref{22}), (\ref{28}) and (\ref{27}). The self-energy components
in Eqs. (\ref{23}-\ref{26}) are zero in the symmetric case. {\em Intrasubband%
} contributions to the scattering rates arise from the terms $\sigma _{1111}$
and $\sigma _{2222}$,\thinspace while {\em intersubband }contributions are
due to the terms $\sigma _{1221}$ and $\sigma _{2112}$. The contributions
coming from the plasmon modes (filled-square lines) are obtained separately
by excluding the continua SPE$_{11}$ and SPE$_{12}$ from the numerical
integrations, whereas contributions coming only from the single-particle
continua are obtained by numerically evaluating Eqs. (\ref{22}) and (\ref{27}%
) only for the region representing each continuum leaving out the plasmon
contributions. Single-particle excitations contribute for all values of wave
vectors $k$. However, neither intra- nor inter-subband plasmon mode
contributes to the scattering rates close to $k_{F1}^{sy}$ or $k_{F2}^{sy}$.
These collective modes provide excitation channels for inelastic relaxation
only above some threshold wavevectors. The intrasubband plasmon mode $(1,1)$%
\thinspace begins to contribute to either $\sigma _{1}(k)$ $\,$or $\sigma
_{2}(k)$ when the wavevector is larger than the same threshold $%
k_{11}^{sy}\simeq $ 1.65$\times $10$^{6}$ cm$^{-1}$. On the other hand, the
contribution coming from the plasmon mode $(1,2)$ has a different threshold
for each scattering rate. This mode begins contributing to $\sigma _{1}(k)$
and $\sigma _{2}(k)$ when the wavevector is larger than the thresholds $%
k_{12}^{sy}\simeq $ 1.25$\times 10^{6}$ cm$^{-1}$ $\,$and $k_{12^{\prime
}}^{sy}\simeq $ 2.0$\times $10$^{6}$ cm$^{-1}$, respectively (notice that $%
k_{F2}^{sy}<k_{F1}^{sy}<k_{12}^{sy}<k_{11}^{sy}<k_{12^{\prime }}^{sy}$).
Obviously, these thresholds depend on the particular choice of sample
parameters. In the present paper, they are smoothed ( instead of being a
very sharp threshold ) because we are considering the impurity induced
constant $\gamma =0.2$ meV in our numerical evaluation. These thresholds
become much sharper for smaller values of $\gamma $ without any other
substantive changes in our numerical results.

Figure 2(b) shows the same results as in Fig. 2(a) but for the asymmetric
bilayer system of Fig. 1(b). In contrast to the symmetric case, where we
were able to separately obtain the inter- and intra-subband plasmon modes
through the roots of $\varepsilon _{inter}=0$ and $\varepsilon _{intra}=0,$
respectively; the coupled plasmon dispersion in the asymmetric system is
obtained directly from the numerical roots of Eq.(\ref{35}) in which the
bare off-diagonal Coulomb interactions $V_{J}$ and $V_{H}$ are now
non-vanishing. The terms in Eq. (\ref{35}) involving $V_{J}$ and $V_{H}$ are
responsible for the mixing between the inter and intra-subband plasmon modes
and for not allowing the contributions coming from the intra- and
intersubband-like plasmon modes $(1,1)\,$ and $(1,2)$ to be picked up
completely separated from each other in the scattering rate. Notice that
dynamically screened Coulomb potential $V_{ijlm}^{s}$ is a full 16$\times $%
16 matrix (for the 2-subband model --- in general, it is an $n^{4}\times
n^{4}$ matrix for an n-subband problem) in the present situation and is
obtained from Eq. (\ref{14}), which involves the dielectric matrix $%
\varepsilon _{ijlm}(q,\omega )$ and the bare Coulomb interactions $%
V_{A},V_{B},V_{C},V_{D},V_{J},$ and $V_{H}$ (all of which are finite in this
strongly coupled asymmetric bilayer structure) \cite{9}. Therefore, both
inelastic scattering rates $\sigma _{1}(k)$ $\,$and $\sigma _{2}(k)$ in the
asymmetric case contain all terms shown in Eqs. (\ref{21}-\ref{28}), which
are finite in this situation. For the sake of clarity and to understand Fig.
2(b) in the same way we did the details in Fig. 2(a), we choose to show
three contributions to the inelastic scattering rates $\sigma _{1}(k)$ $\,$%
and $\sigma _{2}(k)$ in Fig. 2(b) separately: the single-particle
excitations in the continua (i) SPE$_{12}$ (triangles-up) and (ii) SPE$_{11}$%
(diamonds); and (iii) the plasmon mode segment outside these continua
(filled squares). The filled squares in Fig. 2(b) represent contributions
coming from those segments of the plasmon modes which lie outside any
single-particle excitation continua (see Fig.1(b)). Contributions coming
from the plasmon segments lying inside each continuum have been kept in our
numerical work along with the single-particle excitation contributions
because it is essentially numerically impossible to separate the two in this
regime. We should mention that, due to the fact that one is not able to
eliminate the contributions coming from the over-damped plasmon modes lying
inside the Landau continua, the thin lines in Fig. 2(b) serve only the
purpose of qualitative illustration \cite{10}.

\subsection{Coulomb Coupled Bilayers With No Interwell Tunneling}

Now we investigate the two Coulomb coupled identical symmetric quantum wells
of width $W_{1}=W_{2}=$150 \AA\ each with {\em no} interwell tunneling (i.e.
the interwell barrier is taken to be infinity) and with a barrier width of
30 \AA . Here, the Fermi wavevectors in the two wells are of the same value,
i.e. $k_{F0}=k_{F1}=k_{F2}$ $\simeq $ 0.79 $\times 10^{6}$ cm$^{-1}$(or,
equivalently $n_{1}=n_{2}=10^{11}$ cm$^{-2}$). Notice that, the indices 1
and 2 should now be treated as well indices since there is no tunneling
induced bonding-antibonding states. As we mentioned in the Introduction, an
energy degeneracy arises in this case, i.e. $E_{1}=E_{2}$ because the two
wells are identical with no interwell tunneling. If there is no tunneling,
the bare Coulomb potential components $V_{J}=V_{H}=V_{D}=0$ and the
polarizability $\Pi _{12}^{0}=\Pi _{21}^{0}=0$ independent of whether the
bilayer structure is symmetric or not. Besides, for this symmetric
no-tunneling bilayer structure, the bare Coulomb potential $V_{A}=V_{B}$ by
symmetry and the polarizability $\Pi _{11}^{0}=\Pi _{22}^{0}=\Pi _{0}$ due
to the fact that the densities in each well are identical.

According to the Eqs. (\ref{35}-\ref{37}), therefore, the plasmon dispersion
relation should be obtained only from the roots of 
\begin{equation}
\varepsilon _{nt}^{sy}=\left( 1-V_{A}\Pi _{0}\right) ^{2}-V_{C}^{2}\Pi
_{0}^{2}=0.  \label{38}
\end{equation}
Here, the subscript (superscript) $nt$ ($sy)$ stands for no tunneling
(symmetric). As shown in Fig. 3(a), we find four roots of Eq. (\ref{38}).
The solid curves correspond to the in-phase optical, $\omega _{+}(q)$, and
the out-of-phase acoustic, $\omega _{-}(q),$ plasmon modes in the bilayer
structure \cite{11}. These $\omega _{\pm }(q)\,$ modes have been observed 
\cite{12} in multilayer semiconductor systems via inelastic light scattering
spectroscopic experiments. They represent in-phase and out-of-phase
interlayer density fluctuation modes: the out-of-phase acoustic mode, $%
\omega _{-}(q\rightarrow 0)\sim O(q)$ represent densities in the two layers
fluctuating out of phase with a linear wave vector dispersion and the
in-phase optical mode, $\omega _{+}(q\rightarrow 0)\sim \sqrt{N_{e}q}$,$\,$
represent densities in the two layers fluctuating in phase with the usual 2D
plasma dispersion. The dashed lines represent those collective modes which
should be strongly Landau damped by the single-particle excitation continuum
SPE, i.e. the region where $%
\mathop{\rm Im}%
\left\{ \Pi _{0}(q,\omega )\right\} \neq 0$.

Fig. 3(b) shows the same as in Fig. 3(a) but for an asymmetric no-tunneling
situation with the two wells being different, one with a width of $W_{1}=$%
150 \AA \thinspace and the other a width of $W_{2}=$142.4 \AA . In this
case, our no-tunneling bilayer structure is no longer invariant under space
inversion and, consequently, the energy level degeneracy is broken, leading
to the energy $E_{1}<E_{2}$. Besides, the bare Coulomb potential $V_{A}$ is
now different from $V_{B}$. As we discussed before, the two wells now have
different charge densities but we consider the whole system still being in
equilibrium. Furthermore, the Fermi wavevector in the first and second
subband are the same as indicated before, i.e. $k_{F1}^{a}$ and $k_{F2}^{a},$
respectively. Because of the densities in the two wells being different from
each other,$\,$the polarizability $\Pi _{22}^{0}\neq \Pi _{11}^{0}$. The
shadow area in Fig. 3(b) is the single-particle excitation continuum in the
wider quantum well, i.e. the region where $%
\mathop{\rm Im}%
\left\{ \Pi _{11}^{0}(q,\omega )\right\} \neq 0.$ The plasma dispersion
relation is now given by the roots of the Eq. (\ref{36}). Note that all
plasma modes in the zero tunneling system are by definition intrasubband
plasmons in our model where higher subbands are neglected.

\smallskip As shown in Fig. 3(b), we again find four roots of such an
equation and consider that the dashed lines should be strongly Landau damped
modes since they are inside the single particle continua. Furthermore, it
does not make sense, in principle, to define the solid lines in Fig. 3(b) as
pure acoustic or optical plasmon modes because the asymmetry leads to a
difference between the electron densities in each layer. Now, the wider well
has 30\% more electrons than the narrower one and, consequently, the
densities in the two layers are not fluctuating exactly either in phase or
out of phase. The solid lines in Fig. 3(b) are the approximate acoustic- and
optical-like plasmon modes with the strict distinction meaningful only in
the long wavelength limit. Due to the structural asymmetry the acoustic-like
plasmon mode enters the SPE continuum at a smaller wavevector leading to
significant Landau damping of the acoustic plasmon mode by single-particle
excitations in the asymmetric bilayer system. In the single-particle
continuum of the layer 2 (the narrower well) the acoustic-like plasmon mode
is completely suppressed and we find no acoustic-like mode in the $%
\mathop{\rm Im}%
\{\Pi _{22}^{0}(q,\omega )\}\neq 0$ regime. In general the acoustic-like
plasmon mode is found to be much more sensitive to small asymmetry effects
than the optical-like plasmon mode \cite{13}. This is physically reasonable
and should be experimentally tested via inelastic light scattering
experiments.

Now we concentrate on the investigation of the scattering rates $\sigma
_{1}(k)$ $\,$and $\sigma _{2}(k)\,$in the symmetric bilayer structure with
no tunneling. As the bare Coulomb potential $V_{J}=V_{H}=V_{D}=0,\,$it is
straightforward to see that only $V_{1111}^{s}$ and $V_{2222}^{s}$ are
finite in the screened Coulomb interaction matrix $V^{s}$ for a bilayer
structure without any tunneling. Therefore the scattering rates in Eqs. (\ref
{22}-\ref{27}) all vanish by symmetry in this case. The only nonvanishing
terms to be calculated are $\sigma _{1111}$ and $\sigma _{2222}$ in Eqs. (%
\ref{21}) and (\ref{28}), respectively. Furthermore, as we discussed before,
we have the polarizability $\Pi _{11}^{0}=\Pi _{22}^{0}\,=\Pi _{0}$ and the
bare potential $V_{A}=V_{B}$ for identical (i.e. symmetric case) quantum
wells. According to the Eqs. (\ref{V1111})$\,$and (\ref{V2222}), therefore,
the screened Coulomb potential is given by 
\begin{equation}
V_{1111}^{s}=V_{2222}^{s}=\frac{V_{A}\left( 1-V_{A}\Pi ^{0}\right)
+V_{C}^{2}\Pi ^{0}}{\varepsilon _{nt}^{sy}}  \label{1=2}
\end{equation}
in the present situation of a symmetric bilayer system with no interwell
tunneling. In fact, the total inelastic Coulomb scattering $\sigma _{1}(k)$
and $\sigma _{2}(k)$ are identical because the two wells are identical with
the same density. The thick line shown in Fig. 4(a) represents the total
inelastic scattering rate which is equal ($\sigma _{1}(k)=\sigma
_{1111}=\sigma _{2}(k)=\sigma _{2222}$ ) in both subbands, as a function of
the wavevector $k$. To show separately the contributions coming from the
emission of plasmons (squares) and single-particle excitations (diamonds),
we again exclude the region where $%
\mathop{\rm Im}%
\left[ \Pi _{0}(q,\omega )\right] \neq 0$ from the numerical calculations to
obtain the plasmon contribution. Single-particle excitations again
contribute at all values of the wavevector whereas the plasmons begin
contributing to the scattering rate for wavevectors $k$ larger than the
Fermi wavevector $k_{F0}$. There are clearly two thresholds wave vectors in
the plasmon contribution (squares), one at $k=k_{ac}^{sy}\simeq 1.25\times $%
10$^{6}$ cm$^{-1}$ and other at $k=k_{op}^{sy}\simeq 1.65\times $10$^{6}$ cm$%
^{-1}.$ These are the thresholds for the emission of the acoustic and the
optical plasmon, respectively. The substantial difference between Figs. 4(a)
and 2(a) demonstrates the strong effect of tunneling on the inelastic
scattering rates in bilayer structures. This is one of the important new
qualitative results in our paper.

Figure 4(b) shows the same results as in Fig. 4(a) but for the asymmetric
bilayer structure without tunneling. As we discussed before, the asymmetry
leads to $\Pi _{22}^{0}\neq \Pi _{11}^{0}$. Furthermore, the bare Coulomb
potential $V_{A}\neq V_{B}$ and, therefore according to Eqs. (\ref{V1111})$%
\, $and (\ref{V2222}), the screened Coulomb potential $V_{1111}^{s}$ $\neq
V_{2222}^{s}$ in the asymmetric case. In this situation, $\sigma
_{1}(k)=\sigma _{1111}$ (thick solid line) and $\sigma _{2}(k)=\sigma
_{2222} $ (thick dashed line) represent the total inelastic Coulomb
scattering rates in the wider and narrower layer, respectively. They are
different from each other because the two wells have different widths and
densities in the asymmetric situation. Again, we separate the different
contributions (by plasmons and by SPE) by excluding the single-particle
excitation continuum SPE from the numerical calculations to obtain the
plasmon contribution. It is important to point out again that the squares in
Fig. 4(b) represent contributions coming from the emission of undamped
plasmon modes whose frequency $\omega (q)$ lies outside the continuum SPE
(see Fig. 3(b)). There is only one threshold wavevector $k\simeq 1.71\times $%
10$^{6}$ cm$^{-1}$ in the thin solid line (squares) corresponding the
plasmon contribution to $\sigma _{1}(k)$. This threshold is due to the
emission of the optical-like plasmon mode shown in Fig. 3(b). We also find
that, the thin solid line (diamonds) corresponding to the SPE contribution
to $\sigma _{1}(k)$ does not contain any contribution coming from the
acoustic-like plasmon mode. As a matter of fact, there is no contribution to 
$\sigma _{1}(k)$ in Fig. 4(b) coming from the emission of the acoustic-like
plasmon mode at all because the integral in $\sigma _{1}(k)$ does not
contain any segment representing the acoustic-like plasmon mode which is
heavily Landau damped in the asymmetric situation under consideration. On
the other hand, the thin dashed line (squares), corresponding to the plasmon
contribution to $\sigma _{2}(k)$, clearly has two threshold wave vectors $%
k\simeq 1.15\times $10$^{6}$ cm$^{-1}\,$and $k\simeq 1.76\times $10$^{6}$ cm$%
^{-1}$, which characterizes the emission of the acoustic- and optical-like
plasmon mode, respectively. Thus, in the asymmetric case, the acoustic-like
plasmon modes contribute to carrier scattering $\sigma _{2}(k)$ in the
narrower well but not to $\sigma _{1}(k)$ in the wider well by virtue of
strong Landau damping. The difference between Figs. 2(b) and 4(b) represents
the strong effect of tunneling on the second component of the inelastic
scattering rates $\sigma _{2}(k)$ in bilayer asymmetric structures.

\subsection{\protect\smallskip Single Symmetric Quantum Well}

\smallskip We now consider (for the sake of comparison) a single symmetric
GaAs-Al$_{x}$Ga$_{1-x}$As quantum well of width 300 \AA , barrier height 228
meV, and with the same total electron density $N_{e}=2\times $10$^{11}$cm$%
^{-2}$ as used before. These sample parameters lead to the Fermi wave vector
in the first subband $k_{F1}^{\text{single}}\simeq 1.13\times $10$^{6}$cm$%
^{-1}$ with only one subband occupancy. Here, the second subband is empty,
which leads to $\Pi _{22}^{0}=0$. As we discussed before, only the bare
Coulomb potential $V_{A},V_{B}\,,V_{C}$ and $V_{D}$ are finite because $%
V_{J}=V_{H}=0$ in symmetric structure. According to Eqs. (\ref{36}) and (\ref
{37}), therefore, the intra- and inter-subband plasmon modes are obtained
from the roots of the equations 
\[
\varepsilon _{intra}^{\text{single}}=\left( 1-V_{A}\Pi _{11}^{0}\right) =0
\]
$\,$ and 
\[
\varepsilon _{inter}^{\text{single}}=1-V_{D}\left( \Pi _{12}^{0}+\Pi
_{21}^{0}\right) =0,
\]
Taking $\Pi _{22}^{0}=0$ (unoccupied excited subband) in Eqs. (\ref{V1111}),
(\ref{V2222}) and (\ref{V1212s}) we get 
\begin{equation}
V_{1111}^{s}=\frac{V_{A}}{\varepsilon _{intra}^{\text{single}}},
\end{equation}
\begin{equation}
V_{2222}^{s}=\frac{V_{B}\left( 1-V_{A}\Pi _{11}^{0}\right) +V_{C}^{2}\Pi
_{11}^{0}}{\varepsilon _{intra}^{\text{single}}},
\end{equation}
and 
\begin{equation}
V_{1221}^{s}=V_{2112}^{s}=\frac{V_{D}}{\varepsilon _{inter}^{\text{single}}}.
\end{equation}
Again in this case, the screened Coulomb potential $%
V_{1112}^{s}=V_{1121}^{s}=V_{1211}^{s}=V_{2111}^{s}=V_{2221}^{s}=V_{2212}^{s}=V_{2122}^{s}=V_{1222}^{s}=0
$ by symmetry because $V_{J}=V_{H}=0$. Therefore, as we discussed in the
Sec. III.A, the total inelastic scattering rates in the first and the second
subband are given by Eqs. (\ref{gam1}) and (\ref{gam2}), respectively. In
the same way we did for the bilayer structures, we present the scattering
rates $\sigma _{1}(k)$ and $\sigma _{2}(k)$ in Fig. (5) in the thick-solid
and thick-dashed lines, respectively. The symmetric nature of the single
well system enables us to separate out the different contributions to the
scattering rates as discussed before. We find that contributions to $\sigma
_{1}(k)$ come mainly from the emission of both the intrasubband plasmons $%
(1,1)$ and the intrasubband single-particle excitations SPE$_{11}$. The
emission of intersubband excitations turn out to make negligible
contributions to the scattering because of the sufficiently large energy gap
between the two subbands ($\omega _{0}=E_{21}=14.63$ meV). For this
particular choice of the sample parameters, $\sigma _{1111}$ turns out to be
much larger than $\sigma _{1221}$, implying that the carrier relaxation
process in the ground subband is almost entirely via intrasubband
scattering. Another important point in Fig. 5 is that inter- and
intra-subband plasmon modes as well as intra- and inter-subband
single-particle excitations contribute to the total inelastic scattering
rate $\sigma _{2}(k)$ in the second subband. Notice that, in contrast to the
behavior of $\sigma _{1}(k)$, the total inelastic scattering rate in the
second subband $\sigma _{2}(k)$ does not vanish for wavevectors any $k.$
This is due the fact that there is no Fermi surface in the second subband .
This should lead to qualitatively different effects in the measured carrier
injected in the second subband compared with that in the ground subband \cite
{14}. This lifetime, which is inversely proportional to the total inelastic
scattering rate $\sigma _{2}(k),$ should be finite for all finite
wavevectors in the excited empty subband.

\section{\protect\smallskip Conclusions}

We have developed a theory for calculating the inelastic relaxation rate for
Coulomb scattering in coupled bilayer structures in semiconductor double
quantum well systems. We use a many body theory based on a multisubband
generalized GW approximation with the inelastic scattering rate defined by
the magnitude of the imaginary part of the on-shell electron self-energy.
Effects of dynamical screening, mode coupling, and Fermi statistics are
naturally included in our many-body theory. We demonstrate the usefulness of
our theory by obtaining results for general representative two-subband model
systems: Coulomb coupled bilayer GaAs-AlGaAs double quantum well structures
both with and without interwell tunneling and also with and without
interwell asymmetry in the system. Our theory naturally allows for
distinguishing various physical mechanisms contributing to the inelastic
scattering rate: Intra- and inter-subband contributions. We provide a
critical qualitative discussion of these various contributions to scattering
and comment on the effect of interwell tunneling and structural asymmetry in
bilayer quantum wells.

\section{Acknowledgments}

One of us (MRST) is supported by {\em FAPESP}, Brazil. The work at the
University of Maryland is supported by US-ONR. GQH acknowledgs {\em CNPq}
from Brazil for partial support.

\begin{figure}[tbp]
\caption{ Plasmon dispersions in two coupled GaAs/Al$_{0.3}$Ga$_{0.7}$As
quantum wells of widhts (a) $W_{1}=W_{2}=150$ \AA\ (symmetric) and (b) $%
W_{1}=150$ \AA\ and $W_{2}=140$ \AA (asymmetric); and separated by a barrier
of width 30 \AA\ . For the symmetric (asymmetric) situation the energy
separation between the two subbands is $\protect\omega_0=1.68$ meV ($\protect%
\omega_0=2.62$ meV). The shadow areas present the single-particle excitation
regions SPE$_{nn^{\prime}}$ where ${\rm Im} \{\Pi^0_{nn^{\prime}}(q,\protect%
\omega)\} \ne 0$. Each structure is shown in the inset where $\protect\phi %
_{1}(z)$ and $\protect\phi _{2}(z)$ are schematically shown by the solid and
dot lines, respectively. }
\end{figure}

\begin{figure}[tbp]
\caption{ Total inelastic Coulomb scattering rate of electrons in our
coupled bilayer (a) symmetric and (b) asymmetric structures. The thick-solid
and thick-dashed lines denote the total scattering rate $\protect\sigma_n(k)$
for $n$=1 and 2, respectively. The symbols represent the different
contributions to the scattering.}
\end{figure}

\begin{figure}[tbp]
\caption{ Plasmon dispersions in two coupled with no interwell tunneling
GaAs/Al$_{0.3}$Ga$_{0.7}$As quantum wells of widths (a) $W_{1}=W_{2}=150$
\AA\ and (b) $W_{1}=150$ \AA\ (symmetric) and $W_{2}=142.4$ (asymmetric)
\AA\ ; and separated by an infinity barrier of width 28.87 \AA\ . The shadow
areas present the single-particle excitation regions SPE$_{11^{\prime}}$
where ${\rm Im} \{\Pi^0_{11^{\prime}} (q,\protect\omega)\} \ne 0$. Each
structure is shown in the inset where $\protect\phi _{1}(z)$ and $\protect%
\phi _{2}(z)$ are schematically shown by the solid and dot lines,
respectively. }
\end{figure}

\begin{figure}[tbp]
\caption{ Total inelastic Coulomb scattering rate of electrons in our
coupled bilayer (a) symmetric and (b) asymmetric structures with no
tunneling. In the part (b), the thick-solid and thick-dashed lines denote
the total scattering rate $\protect\sigma_n(k)$ for $n$=1 and 2,
respectively. The symbols represent the different contributions to the
scattering.}
\end{figure}

\begin{figure}[tbp]
\caption{ Total inelastic Coulomb scattering rate of electrons in a single
quantum well. The thick-solid and thick-dashed lines denote the total
scattering rate $\protect\sigma_n(k)$ for $n$=1 and 2, respectively. The
symbols represent the different contributions to the scattering.}
\end{figure}


\begin{references}
\bibitem{9}  We do not present here the full 16$\times $16 $V_{ijlm}^{s}$ or 
$\varepsilon _{ijlm}$ matrix in the asymmetric situation because it is not
particularly illuminating. In fact, it is much more convenient to evaluate
everything numerically using the matrix inversion shown in Eqs. (\ref{14})
and (\ref{15}).
\bibitem{10}  Notice that, it is extremely difficult, by numerical means, to
exclude contributions from those segments of the damped plasmon modes lying
inside each single-particle continua when we are calculating the
single-particle excitation contributions to the scattering rate.
\end{references}
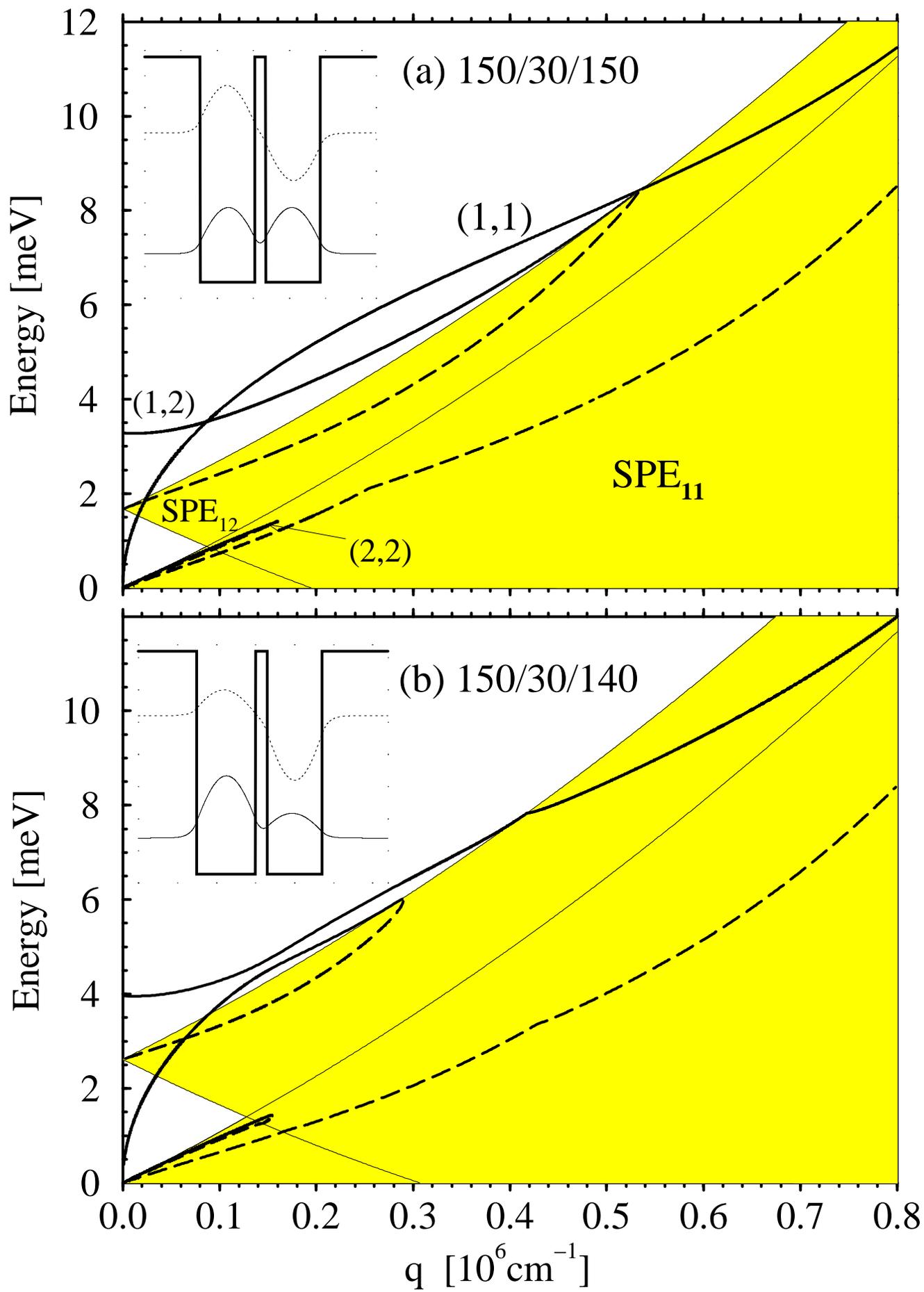
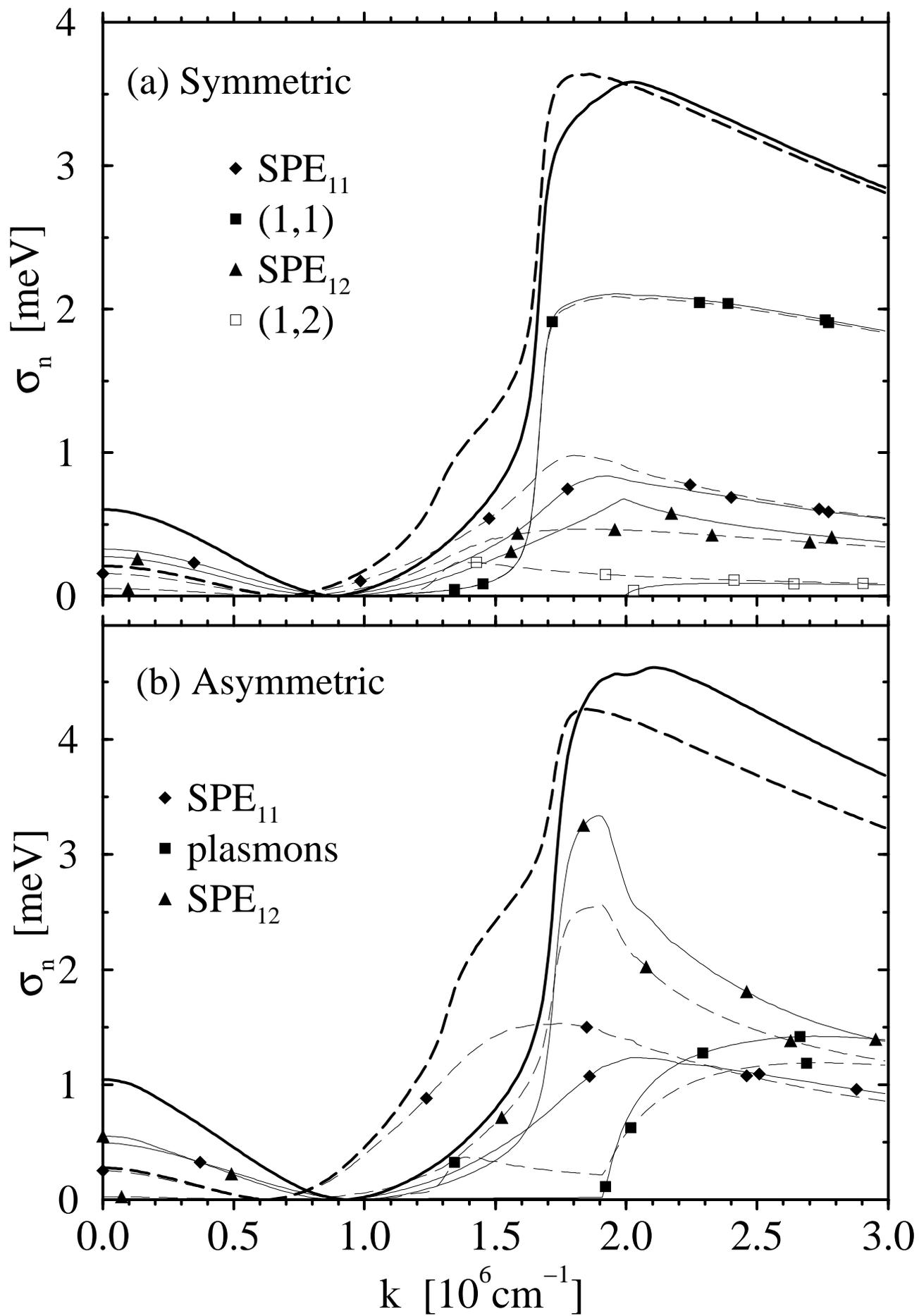
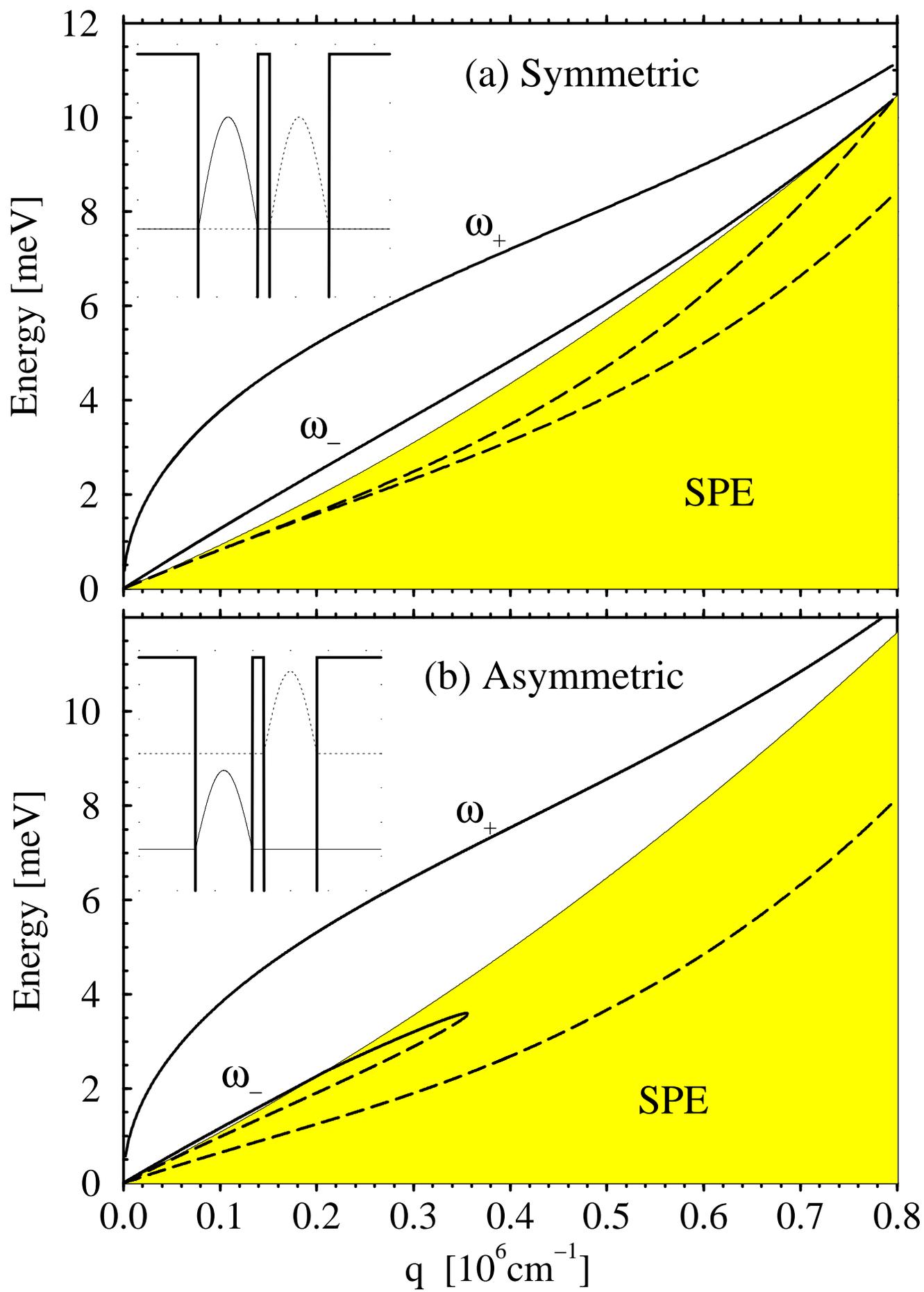
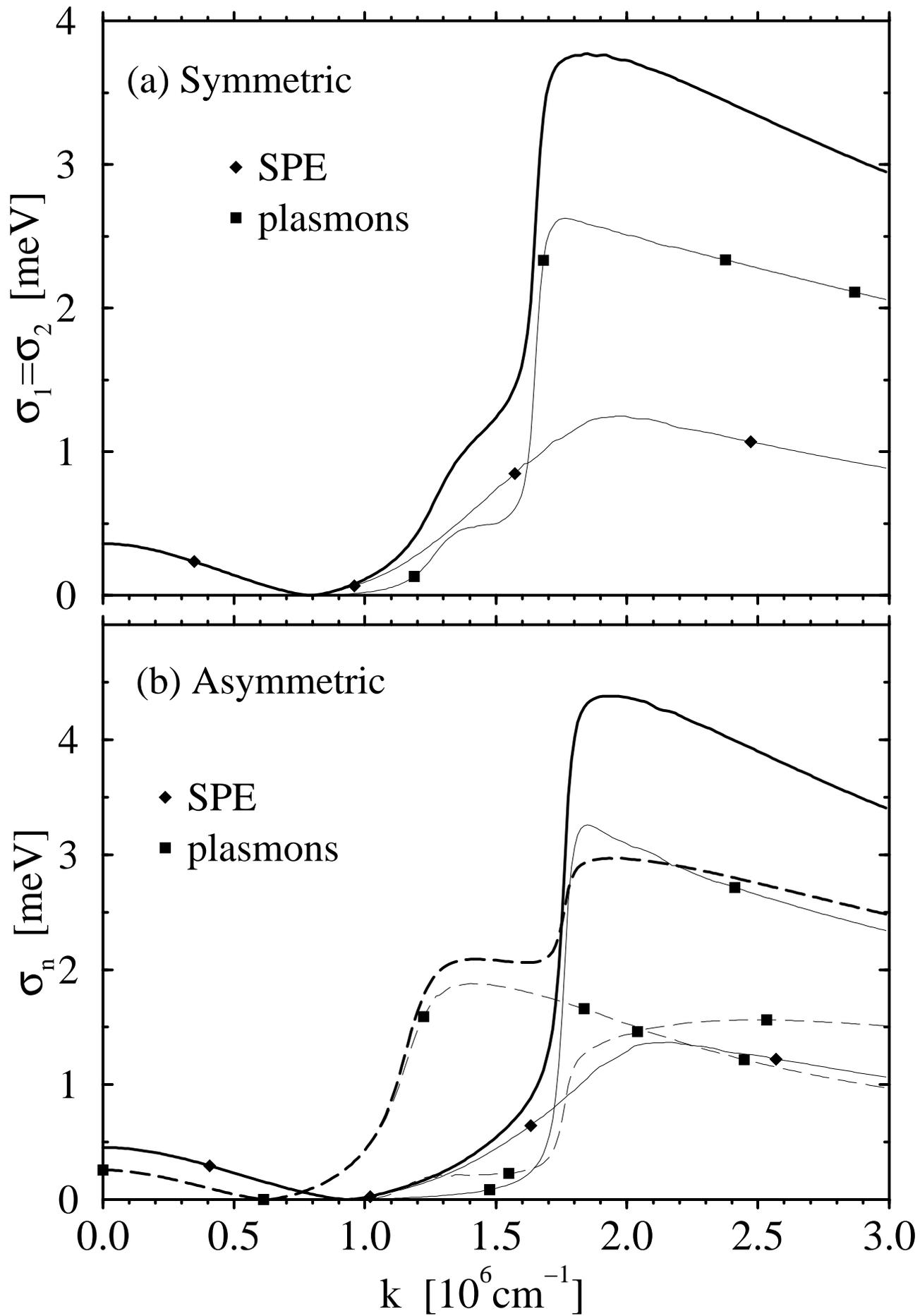
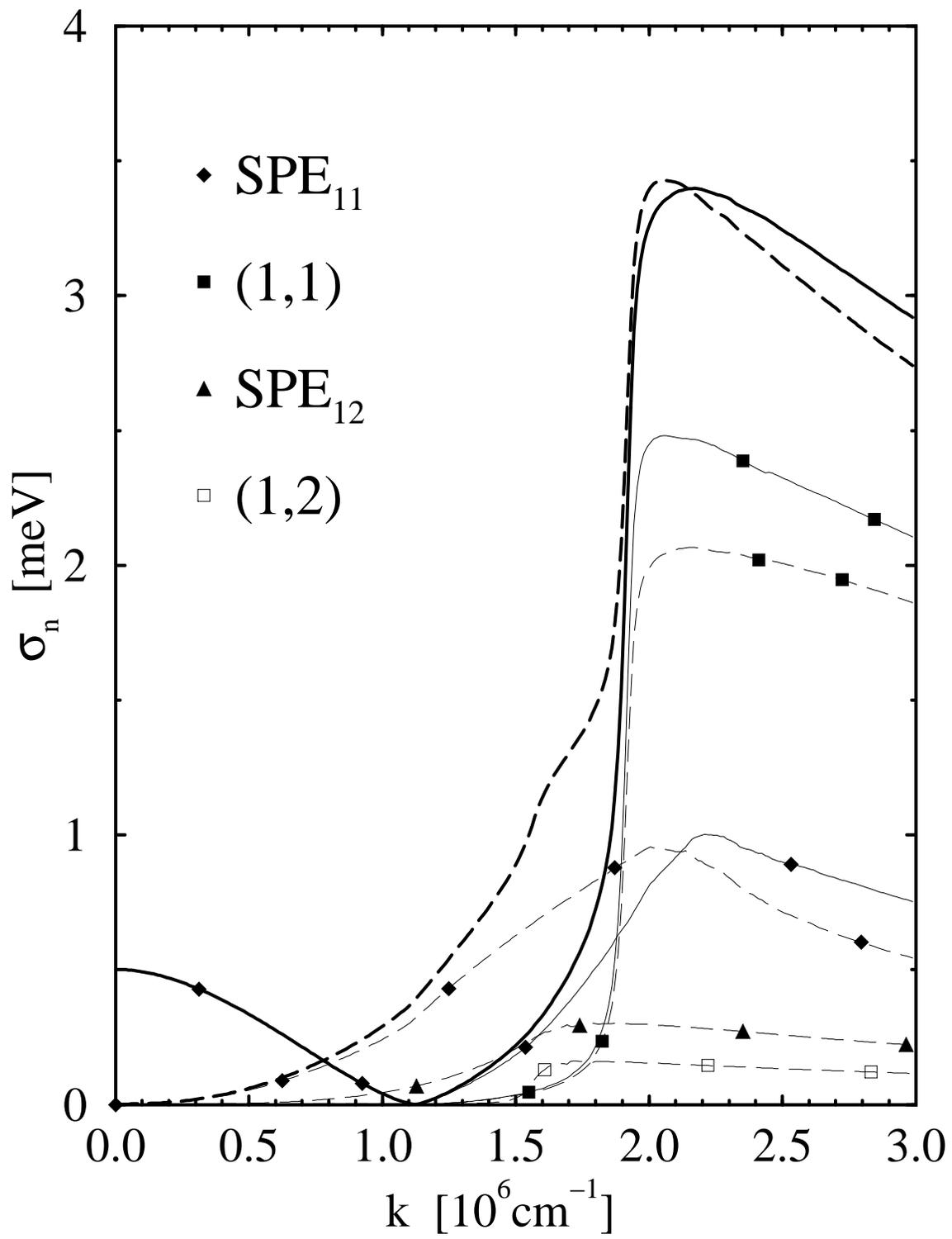
\end{document}